\documentclass[twocolumn]{aastex63}

\usepackage{amsmath}
\usepackage{float}

\newcommand{\gppr}{\stackrel{>}{\scriptstyle \sim}}
\newcommand{\lppr}{\stackrel{<}{\scriptstyle \sim}}

\received{XXX, 2020}
\revised{yyy, 2020}
\accepted{zzz, 2020}
\submitjournal{ApJ}

\shorttitle{Constraining cosmic-ray acceleration in the magnetospheric gaps of Sgr~A$^{*}$}
\shortauthors{Katsoulakos et al.}

\graphicspath{{./}{figures/}}

\begin{document}

\title{Constraining cosmic-ray acceleration in the magnetospheric gaps of Sgr~A$^{*}$}


\correspondingauthor{Grigorios Katsoulakos}
\email{gkats@mpi-hd.mpg.de}

\author[0000-0002-8155-5969]{Grigorios Katsoulakos}
\affiliation{Max-Planck-Institut f\"ur Kernphysik, Saupfercheckweg 1, 69117 Heidelberg, Germany}
\author[0000-0003-1334-2993]{Frank M. Rieger}
\affiliation{Max-Planck-Institut f\"ur Kernphysik, Saupfercheckweg 1, 69117 Heidelberg, Germany}
\affiliation{ZAH, Institut f\"{u}r Theoretische Astrophysik, Universit\"{a}t 
Heidelberg, Philosophenweg 12, D-69120 Heidelberg, Germany}
\author[0000-0002-3778-1432]{Brian Reville}
\affiliation{Max-Planck-Institut f\"ur Kernphysik, Saupfercheckweg 1, 69117 Heidelberg, Germany}

\begin{abstract}
Sagittarius A* (Sgr~A*) is a potential VHE $\gamma$-ray and cosmic-ray 
source. We examine limits to gap-type particle acceleration in the
magnetosphere of Sgr A*, showing that in the current phase of activity
proton acceleration to PeV energies is possible, with injection powers 
into the environment usually limited to several $10^{36}$ erg s$^{-1}$. 
Compton upscattering of ambient soft photons by gap-accelerated 
electrons could yield TeV emission compatible with the detected VHE 
points source. 
We explore the dependency of the results on changes in the accretion rate
showing that higher stages in the past are unlikely to increase the power
output unless the inner accretion flows itself changed its configuration.
\end{abstract}


\section{Introduction} \label{sec:01}

The centre of the Milky Way harbors a supermassive black hole (BH) of mass 
$M_{\rm BH} \simeq 4.3 \pm 0.3 \times 10^6 M_{\odot}$ \citep[e.g.,][]{boe2016,
gil2017}. Its location is coincident with the compact radio source Sgr~A* at a 
distance of $d\simeq 8.2$ kpc \citep{abu2019} that is known to exhibit periods of steady 
and variable, non-thermal emission across the electromagnetic spectrum 
\citep[e.g.,][for review]{gen2010}.

At very high energies (VHE) H.E.S.S. observations of the Galactic Centre 
region have revealed a bright, point-like $\gamma$-ray source spatially 
coincident with Sgr~A*, along with extended ($> 100$ pc) diffuse VHE emission 
correlated with massive gas-rich complexes in the Central Molecular Zone (CMZ)
\citep{HESS2006,HESS2016,HESS2018}. The latter correlation points to a hadronic 
origin of the diffuse emission where the $\gamma$-rays are produced in interactions 
of PeV protons with ambient gas. 
The spatial map of the diffuse VHE emission can thus be used to estimate the radial 
distribution of cosmic-ray (CR) protons within the CMZ. The resultant CR distribution
appears compatible with quasi continuous injection of $>100$~TeV protons from the 
vicinity of Sgr A*, and diffusive propagation for $\sim 10^4$ yr \citep{HESS2016,HESS2018}.
The $\gamma$-ray point source at the Galactic Centre (GC), on the other hand, shows 
a power-law type VHE spectrum (photon index $\simeq 2.1\pm0.1$) from $\sim(0.1-10)$ 
TeV along with evidence for a cut-off \citep[cf.,][]{aha2009,acc2020}, probably 
related to absorption of VHE gamma-rays by the ambient radiation field, and 
exhibits a modest luminosity of $L_{\rm VHE} \sim 10^{35}$ erg s$^{-1}$. 

The current quiescent bolometric luminosity of Sgr~A* is rather low, at a level of 
$L_B\sim 10^{36}$ erg s$^{-1}$ $\sim 2 \times 10^{-9} L_{\rm Edd}$, suggesting that Sgr~A* 
is accreting in a radiatively inefficient mode \citep[e.g.,][]{yua2014}. 
There is X-ray morphological evidence, however, that Sgr A* could have been 
brighter (i.e., temporarily exceeding $10^{38-39}$ erg s$^{-1}$) in the more recent past 
\citep[e.g.,][]{pon2013,zha2015,ter2018}. We note that in the more distant past 
(i.e., a few Myr ago) much higher accretion rates, up to several percent of 
the Eddington value, must have occurred if the Fermi bubbles are indeed caused 
by some former AGN-type jet activity \citep[e.g.,][]{guo2012,yan2012}.

Particle acceleration in the vicinity of the GC black hole has been proposed 
as possible source for the observed VHE radiation and presumed CR injection
\citep[e.g.,][]{aha2005a,aha2005b,lev2011,HESS2016}. In this paper we revisit
the potential of magnetospheric, gap-type particle acceleration for facilitating 
VHE and CR production. This is done by drawing on an advanced steady gap model 
\citep{kat2020} that allows us to incorporate realistic ambient radiation 
fields.

\section{The galactic center BH and vicinity} \label{sec:02}
We assume that the BH in Sgr~A* (horizon scale $r_g= G M_{\rm BH}/c^2$) 
is rotating with angular momentum close to its maximum  $\lesssim G M_{BH}^2/c$. 
The magnetosphere is threaded by a 
magnetic field whose strength is approximately comparable to the equipartition 
value $B_H \simeq 10^6\,\dot{m}^{1/2}$ G, where $\dot{m} =\dot{M}/\dot{M}_{\rm Edd}$ 
denotes the source accretion rate in terms of the Eddington one, $\dot{M}_{\rm Edd}
\simeq 0.1 M_{\odot}$\,yr$^{-1}$. Radio (mm) polarization measurements indicate that 
the current accretion rate close to the black hole is of order, $\dot{M}\sim 
10^{-8}\, M_{\odot}$\,yr$^{-1}$ \citep{bow2018}, while radiative GRMHD models tend 
to favour even lower values, e.g., $\dot{M} \sim 10^{-9}\,M_{\odot}$\,yr$^{-1}$ 
\citep{dra2013}. As noted above, a higher accretion activity might have been 
occurring in the past given the rich gas reservoir present in the GC vicinity 
\citep[cf.,][]{gen2010}. The above values suggest a typical field strength of 
$B_H \sim 100$ G for the present time, roughly compatible with other estimates 
\citep{dex2010,eat2013}.
 
\begin{figure}[t]
\includegraphics[width=0.49\textwidth,height=7.5cm]{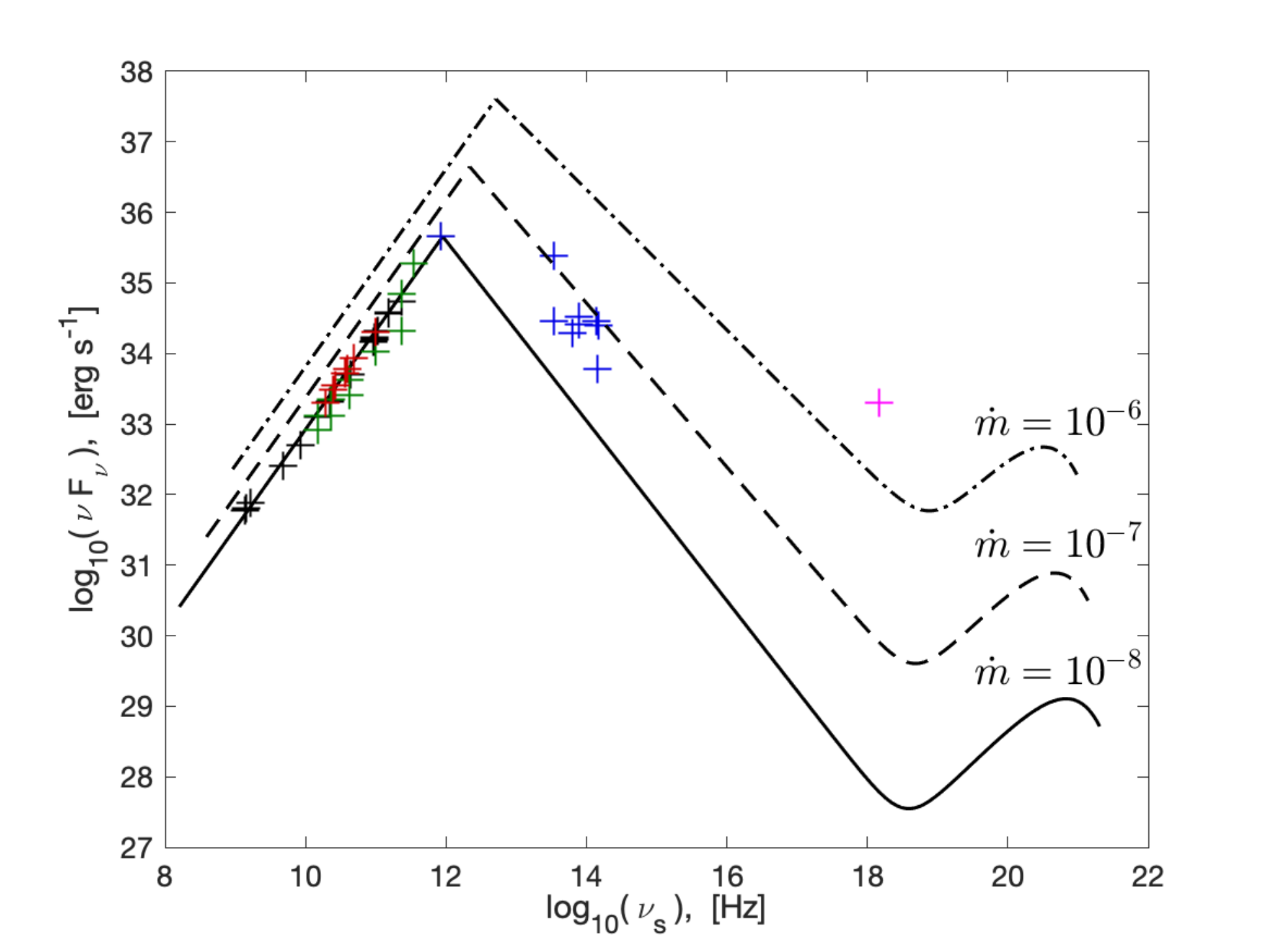}
\caption{Reference ADAF spectra for accretion rates $\dot{m}=10^{-8}$ (solid line), 
$10^{-7}$ (dashed), and $10^{-6}$ (dash-dotted), respectively. A BH mass $M_{\rm BH}
=4.3\times10^{6}M_{\odot}$ has been employed. Data are from 
\citep{ser1997,fal1998,bag2001,hor2002,zha2003,ghe2004,sch2011,bri2015}.}
\label{fig:01}
\end{figure}

In general, the soft photon field from the innermost parts of the accretion flow 
provide a major ingredient for the formation of pair-cascades in the charge-starved 
regions (i.e., gaps) of BH magnetospheres \citep{lev2011,hir2017,kat2020}. 
In the following, we assume the inner accretion flow in Sgr~A* to be hot and
radiatively inefficient (ADAF), though possibly being supplemented by a cool 
gas phase on larger scales \citep[e.g.,][]{yua2014,mur2019}.
We use a simplified ADAF description \citep{mah1997} to characterize the ambient 
soft photon field and its possible variation with accretion rate. 
Figure~\ref{fig:01} shows corresponding disk spectra for three different 
accretion rates, namely, for the current level,
$\dot{m}=10^{-8}$ (solid line), as well as for possible enhanced (past) accretion
periods, i.e., $\dot{m}=10^{-7}$ (dashed line), and, $\dot{m}=10^{-6}$ (dash-dotted
line). To account for the experimental data using the current accretion rate (solid 
line in Fig. \ref{fig:01}), we further assume that 10\% of the 
viscous turbulent energy is attributed to the electrons of the disk, and that the 
peak of the ADAF emission originates from the inner radius of the accretion disk 
located at $r \sim r_s$. The low-energy data are then satisfactorily described, 
while substantial deviations can become apparent toward higher energies. Since 
magnetospheric cascades and gap formation are primarily regulated by the 
low-energy part of the spectrum, i.e., the soft photons around the peak 
\citep{kat2020}, such deviations are not expected to be critical for the present 
purpose. Note that our reference spectra should be considered as a convenient tool 
only, chosen such as to satisfy observational constraints. For detailed ADAF 
modelling of Sgr~A*, we refer to \citet{yua2003}. 

In principle, pair production in a hot accretion flow ($\gamma 
\gamma \rightarrow e^+ e^-$) can lead to charge injection into the BH
magnetosphere. The charge density produced by annihilating MeV photons is 
of the order $n_{\pm}/n_{\rm GJ} \sim 4\times 10^{11} \dot{m}^{7/2}$ 
\citep{lev2011}. Hence, for $\dot{m} \lppr 10^{-4}$ the charge density 
falls short of the Goldreich-Julian (GJ) density $n_{\rm GJ}$, resulting 
in regions of incomplete electric field screening
($\textbf{E}\cdot\textbf{B}\neq 0$), so-called \lq\lq gaps\rq\rq . 
Under these conditions, pair-cascades triggered within these regions
can load the magnetosphere with a significant amount of charges. 

\section{Steady gap acceleration}\label{sec:03}
The parallel electric field component $\mathcal{E}_{||}$ facilitating particle 
acceleration, obeys the generalized Gauss' law \citep[e.g.,][]{kat2020}
\begin{equation}
\nabla\cdot\left(\frac{\mathcal{E}_{||}}{\alpha_{l}}\right)
=4\pi\left(\rho_{e}-\rho_{\rm GJ}\right), \label{eq01}
\end{equation}
where $\rho_{e}$ is the actual charge density, $\rho_{\rm GJ}$ is the GJ 
charge density and $\alpha_{l}$ is the Lapse function \citep{tho1982}. 
The electric field is caused by the difference of the actual charge density 
relative to the GJ value.

Seed electron-positron pairs injected into the gap region (of size $h$) will 
be accelerated along $\mathcal{E}_{||}$, with their energies being limited 
by curvature and inverse Compton (IC) losses. The resultant $\gamma$-rays will 
undergo $\gamma\gamma$-annihilation with soft photons of the accretion disk, 
providing additional pairs to the gap. These secondary leptons will then 
also experience gap acceleration and $\gamma$-ray emission, triggering a third 
generation of leptons, and so on. The ensuing pair-cascade develops until the 
charge density becomes sufficiently large to screen the parallel electric field.

The full gap structure, i.e. the distributions of the parallel electric 
field, the particle energy, the charge and the $\gamma$-ray photon 
densities, can be derived by numerical integration of Gauss' law along 
with the equations of motion and continuity for the pairs, and the 
Boltzmann equation for the $\gamma$-ray photons \citep[e.g.,][]{hir2017,
lev2017,kat2020}. In addition to the BH mass and accretion rate, the
magnetospheric current is a central parameter for steady gap models.
Defined as $J_{0}=\left(\rho_{e}^{-}-\rho_{e}^{+}\right)\,c\,\sqrt{1-1/
\Gamma_{e}^2}$, where $\rho_{e}^{\pm}$ represents the positron/electron 
charge density and $\Gamma_{e}$ the lepton Lorentz factor, the current 
is a constant quantity along magnetic field lines. Since there is 
currently no strong evidence for jet activity in Sgr~A* \citep[though
see also][]{iss2019}, we explore gap solutions for low current values. 
We note that the steady-state solutions for $J_0 \gppr 0.25\,\rho_c c$ in 
our model do not maintain physically consistent values in all quantities 
throughout the gap, and we thus disregard them. 
Here $\rho_{c}=\Omega B_{H}/2\pi c$ is the effective GJ charge 
density. 
In the subsections below, we present gap solutions for different 
accretion regimes following the approach presented in \citet{kat2020}.  

\begin{figure}[htb]
\includegraphics[width=\columnwidth]{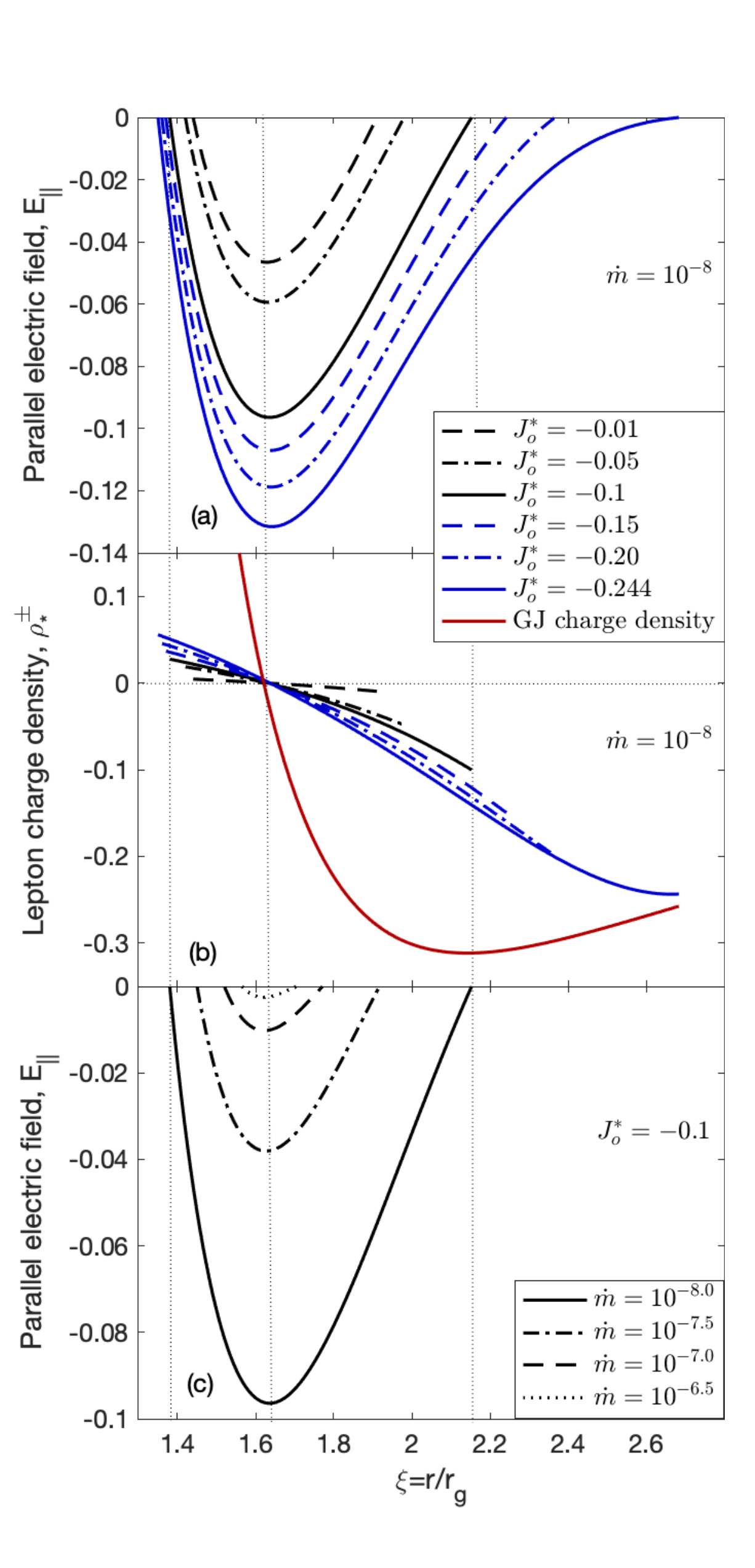}
\caption{Distribution of the (normalized) parallel electric field 
component $\mathcal{E}_{||}^{*r}=\mathcal{E}_{||}^{r}/4\pi\,r_{g}\,
\rho_{c}$, with fixed accretion rate $\dot{m}=10^{-8}$, for different 
current values (upper panel). Normalized 
total charge density, $\rho_{*}^{\pm}=\rho_{\pm}/\rho_{c}$ also for 
fixed $\dot{m}=10^{-8}$. Goldreich-Julian density in red for comparison 
(middle panel).  
Distribution of the (normalized) parallel electric field component 
$\mathcal{E}_{||}^{*r}$ with fixed current $J_0^* = -0.1$, for 
different accretion rates (bottom panel).} \label{fig:02}
\end{figure}

\subsection{Results for the current accretion stage}\label{subsec:03:01}
Figure~(\ref{fig:02},a) summarises the electric field solutions for the present 
accretion rate ($\dot{m}=10^{-8}$) adopting six different values for the 
magnetospheric current, ranging from $J_{o}^{*}=-0.01$ to $J_{o}^{*}=-0.244$, 
where $J_{o}^{*}=J_{o}/(c\,\rho_{c})$. A field line inclination $\theta = 
30^{o}$, a soft photon source size $r_d = 5\,r_g$ and a BH mass $M_{\rm BH} =4.3
\times 10^6 M_{\odot}$ have been adopted throughout.

As can be seen, the gap extension increases as the global current increases. 
Roughly speaking, we obtain gap sizes of the order of $r_g$ for the chosen 
parameters, see Table~\ref{tab_1}. Small current values (e.g., $J_{o}^{*}=
-0.01$) lead to highly under-dense gaps, needing additional charge injection
at the boundaries, while for higher current values (i.e., $J_{o}^{*} =
-0.244$) the GJ charge density at the outer boundary is approached (see 
Fig.~(\ref{fig:02},b)), so that force-free jet formation might occur, 
potentially contributing to the observed emission \cite[e.g.,][]{dav2018}.

We determine the available voltage drop, $\Delta V_{\rm gap}$, by integrating 
$\mathcal{E}_{||}$ along the width of the gap, while the gap power $L_{\rm gap} 
\propto J_0\,\Delta V_{\rm gap}$, is determined by the rate of the lepton energy 
gain multiplied by the number of the particles within the gap. For the 
parameters used here, the gap luminosity typically constitutes only some 
fraction of the available accretion power $L_{\rm acc} = 5.4 \times 10^{37}$ 
erg s$^{-1}$. 
%
\begin{deluxetable}{cccc}[t!]
\vspace{0.5cm}
\tablenum{1}
\tablecaption{Gap properties for the current accretion stage \label{tab_1}}
\tablewidth{0pt}
\tablehead{
\colhead{Global Current} & \colhead{Gap Size} & \colhead{Voltage Drop} 
          & \colhead{Gap Power}  \\
\colhead{$J_{o}^{*}=J_{o}/c\,\rho_{c}$} & \colhead{$h/r_{g}$} & 
\colhead{$\times 10^{15}$ V} & \colhead{$\times 10^{35}$\,erg\,s$^{-1}$} 
}
\startdata
$-0.01$  & $0.47$ & $0.40$ & $0.35$ \\
$-0.05$  & $0.56$ & $0.60$ & $2.38$ \\
$-0.10$  & $0.77$ & $1.27$ & $8.87$ \\
$-0.15$  & $0.87$ & $1.52$ & $15.25$ \\ 
$-0.20$  & $1.01$ & $1.82$ & $24.26$ \\
$-0.244$ & $1.33$ & $2.19$ & $39.22$
\enddata
\tablecomments{Results for a BH mass of $M_{\rm BH} = 4.3 \times 10^6 M_{\odot}$ 
and a fixed accretion rate $\dot{m}=10^{-8}$.}
\end{deluxetable}

In particular, Table~\ref{tab_1} suggests that a power of $L_{\rm gap}\sim 
10^{36}$ erg s$^{-1}$ can be dissipated through the gap, e.g., via particle
acceleration in a voltage difference of $\Delta V_{\rm gap}\sim 10^{15}$ V
($J_{o}^{*}=-0.1$) that could also facilitate PeV CR production. The inferred 
amount of power is comparable to the bolometric luminosity of the GC, indicating 
that gap-type particle acceleration and emission can potentially play a 
dominant role in Sgr~A*. These results depend on the assumption that 
CRs within the gap essentially behave as test particles. The numbers presented
above should thus be viewed as providing firm upper limits on possible CR 
power outputs. While CR injection has often been treated phenomenologically
\citep[e.g.,][]{lev2002,ner2009}, a detailed scenario for CR injection into 
the gap would in principle be needed to quantify the amount of gap power 
carried by CRs.

Our gap solutions yield radiation-limited lepton Lorentz factors $\Gamma_{e}
\lesssim 2\times 10^{8}$. The associated curvature emission peaks at energies
$\epsilon_{\rm cur} =(3/4\pi)(h\,c/r_{g})\,\Gamma_{e}^{3}\lesssim 0.4$ GeV, 
while IC emission reaches up to $\epsilon_{ic} \sim \Gamma_{e} m_{e} c^2 \sim 
10^{2}$ TeV. Absorption of multi-TeV $\gamma$-rays in the ADAF photon fields 
decisively contributes to the cascade development in the gap. The observed
$\gamma$-ray spectrum of Sgr~A* in fact shows a cut-off above $\epsilon_c 
\sim 10$ TeV. Since photons with $\epsilon_c$ preferentially interact with 
soft photons of $\epsilon_s\sim 0.1$ eV, having a spectral luminosity $L_s 
\sim 10^{34}$ erg s$^{-1}$ (Fig.~\ref{fig:01}), the characteristic optical depth 
$\tau_{\gamma\gamma} =  \sigma_{\gamma\gamma}\,n_{s}\, r_{g}$ is of order 
$\tau_{\gamma\gamma} \sim 0.03$, using $\sigma_{\gamma \gamma}\approx 0.2\,
\sigma_{\tau}$ and $n_s=L_s/4\pi \,r_{g}^2\,c\,\epsilon_{s}$. This suggests 
that VHE photons of energy $\epsilon \leq \epsilon_c$ are able to escape from 
the BH vicinity, consistent with observations. Hence, it is possible that at 
the current epoch magnetospheric processes in Sgr~A* may drive both TeV 
$\gamma$-ray as well as PeV CR production.

\subsection{Results for past accretion stages} \label{subsec:03:02}
Changes in the accretion environment will impact on the gap characteristics.
To investigate structural variations of the gap due to possible changes in 
the accretion rate in the past, we also explore higher values, up to $\dot{m}=
10^{-6.5}$, while keeping the current constant ($J_{o}^{*}=-0.1$). The results 
are shown in Fig.~(\ref{fig:02},c) and Table~\ref{tab_2}.

As the ambient soft photon field becomes stronger and cascade formation more 
efficient with higher accretion rates, the gap width essentially decreases 
with increasing accretion rate, i.e., down to $h\sim0.1\,r_{g}$ for $\dot{m} 
= 10^{-6.5}$. As a consequence, the available voltage difference and gap power 
decrease \citep[cf.][]{kat2018}. Thus, despite the fact that the magnetic 
field strength threading the horizon increases, the voltage difference falls, 
$\Delta V_{\rm gap}\lesssim 10^{15}$ V for $\dot{m} \geq 10^{-7.5}$, 
diminishing the potential for PeV CR production. Similarly, achievable 
electron Lorentz factors are reduced to $\Gamma_{e}\sim 6\times 10^6$ for 
$\dot{m}= 10^{-6.5}$. Table~\ref{tab_2} suggests an approximate dependence 
$\Delta V_{\rm gap} \propto \dot{m}^{-1}$ and $L_{\rm gap}\propto 
\dot{m}^{-0.6}$ over the range considered. 
\vspace{-0.5cm}

\begin{deluxetable}{cccc}[htb]
\vspace{0.5cm}
\tablenum{2}
\tablecaption{Gap properties for different accretion rates}\label{tab_2}
\tablewidth{0pt}
\tablehead{
\colhead{Accretion Rate} & \colhead{Gap Size} & \colhead{Voltage Drop}  & 
\colhead{Gap power}  \\
\colhead{$\dot{m}=\dot{M}/\dot{M}_{Edd}$} & \colhead{$h/r_{g}$} 
& \colhead{$\times 10^{15}$ V}
& \colhead{$\times 10^{35}$\,erg\,s$^{-1}$}}
\startdata
$10^{-8.0}$ & $0.77$ & $1.27$  & $8.87$ \\ 
$10^{-7.5}$ & $0.47$ & $0.57$  & $7.08$ \\ 
$10^{-7.0}$ & $0.25$ & $0.15$  & $3.37$ \\ 
$10^{-6.5}$ & $0.14$ & $0.036$ & $1.31$ 
\enddata
\tablecomments{Results for a fixed global current $J_{o}^{*}=-0.1$.}
\end{deluxetable}

\vspace{-1.0cm}
\section{Conclusions}
The above results suggest that at the present accretion stage, the BH 
in Sgr A* is in theory a rather effective electron and CR accelerator. As 
such, IC-upscattering in Sgr~A* could in principle contribute to the GC 
point-source seen by current VHE instruments \citep[e.g.,][]{aha2009,
arc2016,acc2020}. While full radiative modeling is required, a spectral 
cut-off above $\sim10$ TeV, related absorption of VHE gamma-rays by 
the ambient disk photon field is likely to remain a persistent feature 
of gap-related VHE emission. With its superior resolution, the upcoming 
CTA will soon make it possible to probe deeper into the true nature of 
the GC VHE source \citep{cta2019}. Complementary EHT observations are
likely to shed further light on the innermost accretion flow in Sgr~A*
\citep{eht2019}.

The accessible voltage differences in the BH magnetosphere of Sgr~A* can exceed 
$\sim10^{15}$ V, allowing for PeV CR production. Our 
results suggest the power for quasi-continuous CR injection into the GC region 
to be limited to several $10^{36}$ erg s$^{-1}$. If the diffuse VHE emission in 
the CMZ were to be related to the GC BH \citep{HESS2016}, this would thus 
constrain the (average) spatial diffusion coefficient within the CMZ to 
$D \lesssim 10^{29}$ cm$^2$/s for $> 10$ TeV protons. While restrictive, 
this would still be compatible with empirical diffusion models 
suggesting $D \simeq 5\times 10^{28} (E/10~\mathrm{TeV})^{1/3}$ cm$^2$ s$^{-1}$
\citep[e.g.,][]{str2007,fuj2017}. Progress in characterizing the 
turbulence field structures that ultimately determine the CR transport 
properties within the CMZ will help to better assess this. In principle,
magnetospheric gaps are likely to produce rather hard and narrow particle 
distributions. While propagation effects and variable accretion rate will modify
any source spectra when viewed on larger spatial scales, a signature of CR 
acceleration from the gaps might reveal itself through a harder spectral component, 
observable closer in.

Though the GC BH could be a CR PeVatron, no significant contribution to 
the observed Galactic CR spectrum is expected under normal conditions. 
In fact, provided the disk remains ADAF-type, the gap power and potential 
do not increase considering higher accretion stages in the past, as the 
gap extension becomes smaller with higher accretion rates. An exception 
to this could be possible however, for extreme states in the past in which 
the inner accretion flow changed its configuration. This might have 
occurred during the GC phase associated with the generation of the Fermi 
bubbles $\sim 1-10$ Myr ago \citep[e.g.,][]{guo2012,fuj2017,jau2018}, 
and deserves further investigation. 

While the results shown here are based on a simplified disk and magnetic 
field model, we expect them to be quite generic for quasi-steady gap 
models. Exploring varying disk emission and the characteristics of 
non-steady gaps where the lepton multiplicity could potentially 
exceed one \cite[e.g.,][]{lev2018}, is a goal of future work.

\acknowledgments
We are grateful to Felix Aharonian, John Kirk and Heino Falcke 
for comments and discussions. F.M.R. acknowledges funding by a DFG 
Heisenberg Fellowship under RI 1187/6-1.

\bibliographystyle{aasjournal}
\bibliography{gap}

\end{document}